\documentclass{osa-article}

\journal{oe}

\articletype{Research Article}

\begin{document}

\title{Image watermarking and fusion based on Fourier single-pixel imaging with weighed light source}

\author{ZHIYUAN YE,\authormark{1} PANGHE QIU,\authormark{2} HAIBO WANG,\authormark{1} and JUN XIONG,\authormark{1,*}, KAIGE WANG\authormark{1}}

\address{\authormark{1}Department of Physics, Applied Optics Beijing Area Major Laboratory,
\\ Beijing Normal University, Beijing 100875, China\\
\authormark{2}Department of Physics,Capital Normal University, Beijing 100048, China}

\email{\authormark{*}junxiong@bnu.edu.cn} 

\begin{abstract}
In previous single-pixel imaging systems, the light source was generally idle with respect to time. Here, we propose a novel image fusion and visible watermarking scheme based on Fourier single-pixel imaging (FSPI) with a multiplexed time-varying (TV) signal, which is generated by the watermark pattern hidden in the light source. We call this scheme as TV-FSPI. With TV-FSPI, we can realize high-quality visible image watermarking, encrypted image watermarking and full-color visible image watermarking. We also discuss the extension to invisible watermarking based on TV-FSPI. Furthermore, we don't have to recode illumination patterns, because TV-FSPI can be extended to existing mainstream illumination patterns, such as random illumination mode and Hadamard illumination mode. Thus TV-FSPI has the potential to be used in single-pixel broadcasting system and multi-spectral single-pixel imaging system.
\end{abstract}

\section{Introduction}
Single-pixel imaging (SPI), as an extension of computational ghost imaging (CGI) \cite{Shapiro2008,Bromberg2008}, has been developed rapidly in the past decade \cite{Sun2013,Sun2016,Soldevila2018,Hu2019,Liu2019,Radwell2014,Stantchev2016}. Active SPI requires a series of time-varying (TV) illumination patterns to encode the two-dimensional spatial information of the object into a one-dimensional TV light intensity sequence, which is usually acquired synchronously by a single-pixel detector (SPD) \cite{Edgar2019}. In order to generate TV illumination patterns quickly, spatial light modulators or digital micromirror devices are often used in SPI. The imaging mechanism of SPI has been applied to many areas, such as three-dimensional imaging  \cite{Sun2013,Sun2016}, wave-front phase imaging \cite{Soldevila2018,Hu2019,Liu2019}, non-visible band imaging \cite{Radwell2014,Stantchev2016}.

The TV sequence of light intensity acquired from the SPD can be decoded to obtain an image of the object. With the development of SPI, researchers have proposed various decoding methods \cite{Ferri2010,Sun2012,Khamoushi2015,Li2013,Zhang2014,Gong2015,Duarte2008}, such as traditional correlation operations \cite{Ferri2010,Sun2012,Khamoushi2015}, correspondence operations \cite{Li2013}, pseudo-inverse operations \cite{Zhang2014,Gong2015}, compressed sensing \cite{Duarte2008}, etc. In 2014, Zhang et al. presented a novel Fourier single-pixel imaging (FSPI) technique, in which the phase-shifting sinusoid structured illumination and the inverse Fast Fourier transform (IFFT) were employed to obtain the high-quanlity images \cite{Zhang2015}. Subsequently, a series of improved expansion schemes based on FSPI were reported \cite{Zibang2018,Huang2018,Jiang2018,Czajkowski2018}.

Digital visible watermarking or image fusion technology has been widely used in the field of public information security such as copyright protection \cite{Li1995,Cox2002,Hu2005,Thodi2007,Xu2016,Yang2018}. In SPI, many image fusion or watermarking schemes \cite{Chen2014,Li2012,Sunm2013,Dongfeng2017,Liansheng2018,Zhang2019} have been proposed. Some of them \cite{Chen2014,Li2012,Sunm2013} multiplexed spatial information of the object to achieve image fusion or encryption, but their imaging quality and efficiency need to be improved and their schemes do not have high concealment. Other schemes \cite{Dongfeng2017,Liansheng2018,Zhang2019} often need to re-encode the light patterns and may require additional complex operations to extract information, limiting their usefulness in public detection \cite{Zhong2018} and information security.

In active SPI or CGI, the fluctuations in ambient illumination level often lead to degradation in the signal-to-noise ratio (SNR) of the reconstructed image \cite{SunM2016}. Therefore, the light source used in SPI is usually a stable light source. In other words, the light source in SPI was generally idle with respect to time. Photon, as a carrier of information coding, is always hoped to expand its degrees of freedom, such as orbital angular momentum. In our view, however, TV light source can be regarded as a new degree of freedom for unusual multiplexing in SPI.

In this study, we propose a novel FSPI-based high-quality visualization of image fusion and watermarking scheme that multiplexes TV signals hidden in the light source. The scheme is named as TV-FSPI. We calculate the Fourier coefficients of the watermark image in advance and load it into the sinusoid structured illumination pattern. Note that the arrangement of the illumination patterns does not change, except that the intensity of each illumination pattern changes over time. Moreover, the TV signals are hidden in the light source of the transmitting end and thus this scheme has high concealment at the receiving end. Principles, simulations and experiments demonstrate the effectiveness of this scheme. We also discuss the scheme of invisible watermarking based on TV-FSPI.

\section{Principles}

According to four-step phase-shifting approach \cite{Zhang2015}, the sinusoid illumination patterns $P_{\varphi}$ can be generated by
\begin{equation}
P_{\varphi}(x,y;f_{x},f_{y})=a+b\cos(2\pi f_{x}x+2\pi f_{y}y+\varphi),\label{eq:1}
\end{equation}
where $(x,y)$ represents the 2-D Cartesian coordinates in the scene; $a$ and $b$ are real constants that represent the direct current and the contrast of the illumination patterns (i.e.,$a=b=255/2$ ); $\varphi$ represents the initial phase. Every illumination pattern is specified with its spatial frequency $(f_{x},f_{y})$  and initial phase $\varphi$. This sinusoid pattern is projected onto the scene and a SPD acquires the intensity value $I_{\varphi}(f_{x},f_{y})$  of the reflected light simultaneously. This physical process can be expressed as
\begin{equation}
I_{\varphi}(f_{x},f_{y})=\iint_{\Omega} R(x,y)P_{\varphi}(x,y;f_{x},f_{y})dxdy,\label{eq:2}
\end{equation}
where $\Omega$  represents the illuminated area, and $R$ is the distribution of the scene to be imaged. This four-step phase-shifting approach allows each complex Fourier coefficient to be acquired by every four corresponding illumination patterns with different phase $\varphi$  (i.e.,$P_{0}$, $P_{\pi/2}$,  $P_{\pi}$, and $P_{3\pi/2}$). In this way, we can obtain each complex Fourier coefficient $C(f_{x},f_{y})$ as
\begin{equation}
C(f_{x},f_{y})=[I_{0}(f_{x},f_{y})-I_{\pi}(f_{x},f_{y})]+j\cdot[I_{\pi/2}(f_{x},f_{y})-I_{3\pi/2}(f_{x},f_{y})],\label{eq:3}
\end{equation}
where $j$ denotes the imaginary unit. The image of the scene can be recovered using IFFT,  that is
\begin{equation}
\hat{R}=\textrm{IFFT}(C),\label{eq:4}
\end{equation}
where $\hat{R}$  represents the estimated solution of $R$. In fact, a large amount of noise is introduced during the actual measurement process, and this differential measurement can greatly suppress the interference of noise.

\begin{figure}[!ht]
\centering
\includegraphics[width=13cm]{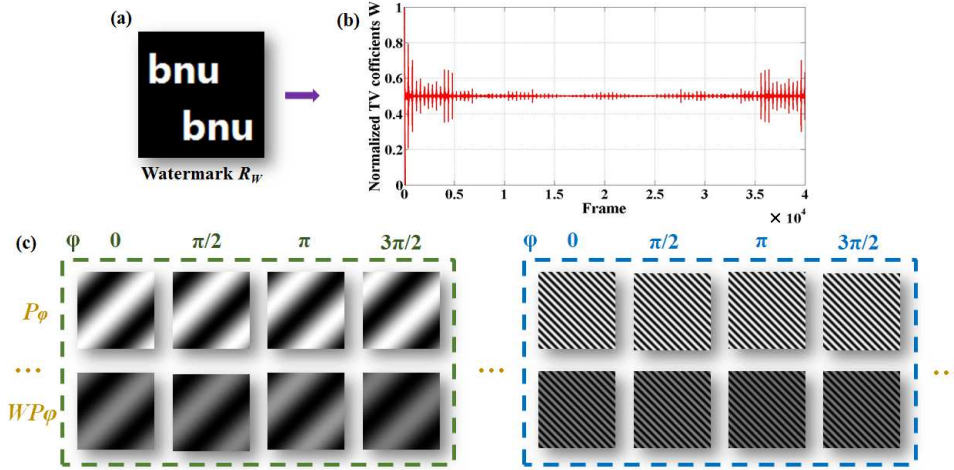}
\caption{Schematic diagram of TV-FSPI scheme: (a) the watermark pattern $R_{W}$  (100 $\times$ 100 pixels); (b) the four-step normalized coefficients $W$ of the watermark pattern; (c) the four-step illumination patterns with the TV signals.}
\end{figure}

The above is the basic principle of the FSPI scheme. In this paper, we hide the TV signal in the illumination patterns to realize optical image watermarking and fusion on the basis of FSPI. Fig.1 shows the schematic diagram of TV-FSPI scheme. We calculate the four-step coefficients $W_{\varphi}(f_{x},f_{y})$   of the watermark pattern  $R_{W}$  by computer in advance, which are normalized and loaded into the corresponding illumination patterns as a multiplicative TV sequence. Similar to Eqs. (\ref{eq:2}-\ref{eq:4}), the following equations can be obtained
\begin{equation}
W_{\varphi}(f_{x},f_{y})=\iint_{\Omega} R_{W}(x,y)P_{\varphi}(x,y;f_{x},f_{y})dxdy,\label{eq:5}
\end{equation}
\begin{equation}
C_{W}(f_{x},f_{y})=[W_{0}(f_{x},f_{y})-W_{\pi}(f_{x},f_{y})]+j\cdot [W_{\pi/2}(f_{x},f_{y})-W_{3\pi/2}(f_{x},f_{y})],\label{eq:6}
\end{equation}
\begin{equation}
\hat{R}_{W}=\textrm{IFFT}(C_{W}).\label{eq:7}
\end{equation}
When we modulate the relative intensity of the sinusoid illumination patterns with a watermark, the physical process depicted in Eqs. (\ref{eq:2})  and  (\ref{eq:3}) can be rewritten as
\begin{equation}
I'_{\varphi}(f_{x},f_{y})=\iint_{\Omega} R(x,y)[W_{\varphi}(f_{x},f_{y}) \cdot P_{\varphi}(x,y;f_{x},f_{y})]dxdy,\label{eq:8}
\end{equation}
with the corresponding complex Fourier coefficient being
\begin{equation}
\begin{split}
C'(f_{x},f_{y})=&[I_{0}(f_{x},f_{y})W_{0}(f_{x},f_{y})-I_{\pi}(f_{x},f_{y})W_{\pi}(f_{x},f_{y})] \\
&+j\cdot  [I_{\pi/2}(f_{x},f_{y})W_{\pi/2}(f_{x},f_{y})-I_{3\pi/2}(f_{x},f_{y})W_{3\pi/2}(f_{x},f_{y})]\\
=&\frac{1}{2}[(W_{0}+W_{\pi}+W_{\pi/2}+W_{3\pi/2})C(f_{x},f_{y})+(I_{0}+I_{\pi}+I_{\pi/2}+I_{3\pi/2})C_{W}(f_{x},f_{y})]\\
&-\frac{1}{2}[(W_{0}-W_{\pi})(I_{\pi/2}+I_{3\pi/2})+(I_{0}-I_{\pi})(W_{\pi/2}+W_{3\pi/2})]\\
&-j\cdot \frac{1}{2}[(W_{\pi/2}-W_{3\pi/2})(I_{0}+I_{\pi})+(W_{0}+W_{\pi})(I_{\pi/2}-I_{3\pi/2})].\label{eq:9}
\end{split}
\end{equation}
Here, we notice a close intrinsic link between the four-step sinusoid illumination patterns from Eq. (\ref{eq:1}), which satisfies
\begin{equation}
P_{0}(x,y;f_{x},f_{y})+P_{\pi}(x,y;f_{x},f_{y})=P_{\pi/2}(x,y;f_{x},f_{y})+P_{3\pi/2}(x,y;f_{x},f_{y})=2a.\label{eq:10}
\end{equation}
We also define two new parameters $K_{1}$  and $K_{2}$,
\begin{equation}
K_{1}=I_{0}+I_{\pi}=I_{\pi/2}+I_{3\pi/2}=2a \iint_{\Omega} R(x,y)dxdy,\label{eq:11}
\end{equation}
\begin{equation}
K_{2}=W_{0}+W_{\pi}=W_{\pi/2}+W_{3\pi/2}=2a \iint_{\Omega} R_{W}(x,y)dxdy,\label{eq:12}
\end{equation}
which represent the sum of the relative amounts of information in the corresponding images, respectively. Of course, in actual situation, there would be noise disturbance. We will discuss the effect of noise on imaging in the next section.

Substituting Eqs. (\ref{eq:11}) and (\ref{eq:12}) into Eq. (\ref{eq:9}) yields
\begin{equation}
C'(f_{x},f_{y})=\frac{1}{2}[K_{2}C(f_{x},f_{y})+K_{1}C_{W}(f_{x},f_{y})].\label{eq:13}
\end{equation}

Thus, the image of the scene illuminated by the light source with a hidden watermark pattern can be reconstructed by
\begin{equation}
\hat{R}'=\textrm{IFFT}(C')=\frac{K_{2}}{2}\hat{R}+\frac{K_{1}}{2}R_{W}.\label{eq:14}
\end{equation}

Eq.(\ref{eq:14}) implies an interesting phenomenon, that is, when the multiplicative TV signal corresponding to the watermark pattern is hidden in the illumination patterns, the final image will result a weighted superposition of two images, which is quite different from traditional image fusion schemes and seems to be contrary to common sense. Moreover, the weighting factor satisfies an interesting inverse relationship. For example, the larger the total amount of information of the watermark pattern $K_{2}$, the weaker the relative intensity of the watermark pattern in the fused image. This graceful rule will be useful for adjusting the relative intensity of the watermark pattern in the fused pattern, which also implies the high flexibility of TV-FSPI scheme. To facilitate the description of the relative relationship between $K_{1}$  and  $K_{2}$, we introduce a new coefficient $Q$ as
\begin{equation}
Q=\frac{K_{1}}{K_{2}}.\label{eq:15}
\end{equation}

The convolution theorem states that the Fourier transform of the convolution of two images is the product of their Fourier transforms. The conclusions of this study seem to be contrary to the theorem, but it is not. FSPI technique does not directly measure the spectral information of an object, but indirectly through four steps. Therefore, the multiplicative TV signal in this study is loaded in the illumination patterns for indirect measurements, rather than being directly loaded into the spectrum.

To quantitatively describe the performance of this visible watermarking technique, we introduce two mainstream indicators for evaluating imaging quality, peak signal-to-noise ratio (PSNR) and structural similarity index (SSIM) \cite{Alain2010}. The formula for calculating PSNR is as follows
\begin{equation}
PSNR = 10{\log _{_{10}}}\left[{\frac{{{{\left( {{2^n} - 1} \right)}^2}}}{{MSE}}} \right],\label{eq:16}
\end{equation}
where $n$ represents the gray level of the image, which is generally regarded as 8 bits;  MSE is the mean square error, which can be expressed as
\begin{equation}
MSE = \frac{1}{{lw}}\sum\limits_{i = 1}^l {\sum\limits_{j = 1}^w {[x(i,j) - y(i,j)} {]^2}} ,\label{eq:17}
\end{equation}
where $l$ and $w$ denote the length and width of the input image $x$ or $y$, respectively.
\begin{equation}
SSIM = \frac{{\left( {2{u_x}{u_y} + {c_1}} \right)\left( {2{\sigma _{xy}} + {c_2}} \right)}}{{\left( {u_x^2 + u_y^2 + {c_1}} \right)\left( {\sigma _x^2 + \sigma _y^2 + {c_2}} \right)}},\label{eq:18}
\end{equation}
where $u_{x}$ and $u_{y}$ represent the average value of input iamges $x$ and $y$, respectively; $\sigma _x^2$ and $\sigma _y^2$ are the variance of input images $x$ and $y$, respectively; $\sigma _{xy}$ denotes the covariance of input images $x$ and $y$; $c_{1}$ and $c_{2}$ are constants used to maintain stability.

\section{Numerical simulations and experiments}
\begin{figure}[!ht]
\centering
\includegraphics[width=8cm]{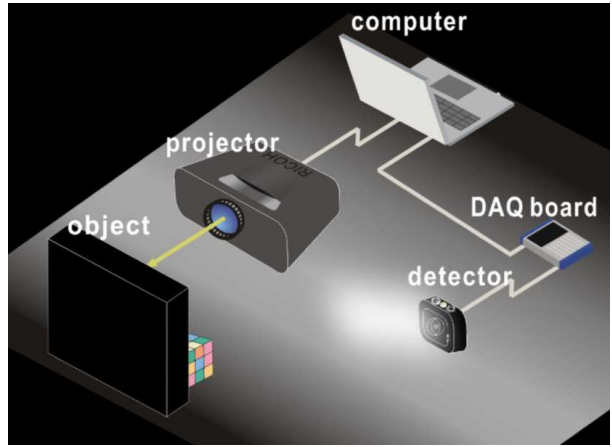}
\caption{Schematic diagram of the experimental setup.}
\end{figure}
In this section, we present a series of numerical simulations and optical experiments to verify the theory in Section 2. Fig. 2 shows the schematic diagram of the experimental setup. To simplify the experiment, we choose a commercial digital projector (Sony, VPL-EX146) to produce structured illumination. At the same time, the TV signal is generated in advance by the watermark pattern in the computer, normalized, and then hidden in the intrinsic illumination sequence, that is, each illumination pattern corresponds to a weighting coefficient as a whole. Since the projector is a combination of a light source and a complex modulation system, it can achieve stable grayscale modulation, so we do not add an additional intensity modulator to generate a TV light source. However, it is clear that the modulation mode in this study is a still sinusoidal illumination pattern and has not been re-encoded. As a proof-of-principle experiment, the projector's modulation speed is set at 10 frames per second. In the acquisition systems, the light intensity of each illumination pattern projected onto the target is detected by the SPD (Thorlabs, PDA100A2) and transmitted to the host computer via a data acquisition card (National Instruments, PCIe-6251) and its accessories (National Instruments, BNC-2110). The sampling rate of the data acquisition card is set to 120KS/s and the amplification gain of the SPD is set at 40 dB. Finally, the collected signals are used for data processing and image reconstruction using Eqs.(\ref{eq:3}) and (\ref{eq:4}) on the host computer. The average amplitude of voltage detected by the SPD in the case of no light is 40 mV, while the average amplitude of voltage can reach 2 V when the projector starts working. Therefore, the SNR of this system can be calculated as 34 dB ( $SNR$=10log$_{10}[(v_{s}/v_{n})^{2}$], where  $v_{s}$ is the average amplitude of the signal, and  $v_{n}$ is the average amplitude of the noise). Due to the differential detection in the FSPI technique, the background term can be well eliminated \cite{Zhang2015}, which also paves the way for visualized high-quality image watermarking and fusion in our study.

\subsection{Visible watermarking}
\begin{figure}[!ht]
\centering
\includegraphics[width=13cm]{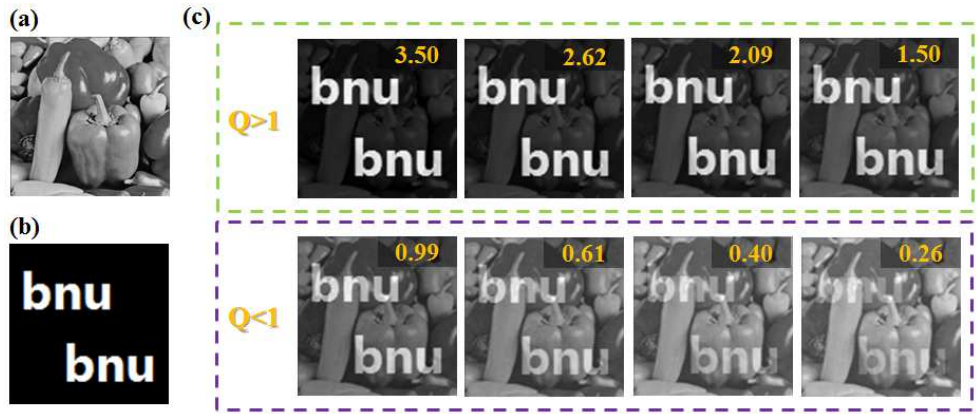}
\caption{Numerical simulation results of FSPI-based visible watermarking: (a) host image of "peppers"; (b) watermark image of "bnu" logo; (c) FSPI-based watermarked images under different values of $Q$ (the yellow number in the upper right corner of the images)}
\end{figure}
We first perform the simulations of image watermarking based on TV-FSPI. The host image used in the simulation is the "peppers" image of size $100\times100$ pixels (Fig. 3(a)), and the watermark image is a "bnu" logo of size $100\times100$ pixels (Fig. 3(b)). Fig. 3(c) shows the simulation results of TV-FSPI-based watermarked images under different values of $Q$. The value of $Q$ can be easily calculated in the numerical simulations. Here, we adjust the relative intensity of the watermark by changing the direct component (DC) of the watermark pattern, that is, adding a constant DC to all pixels' values of the watermark before it's converted into TV signals (see Figs. 4(b1)-4(d1) or 4(e1)-4(g1)). It can be seen from the simulations that as the value of $Q$ increases (i.e., the DC of the watermark is smaller), the relative intensity of the watermark pattern gradually increases. This phenomenon is basically consistent with the theoretical prediction, which satisfies an inverse relationship. It should be emphasized that the imaging result is additive although the TV signal is multiplicative. When the value of $Q$ is 3.5, the DC is equal to 0 at this time, which means that the black background is 0, but the host image can still be seen in the corresponding area in the simulation. This phenomenon also indicates the correctness of the principle and the particularity of the way image fused. Moreover, this property can be used to flexibly adjust the visibility of the watermark in active SPI.

\begin{figure}[!ht]
\centering
\includegraphics[width=13cm]{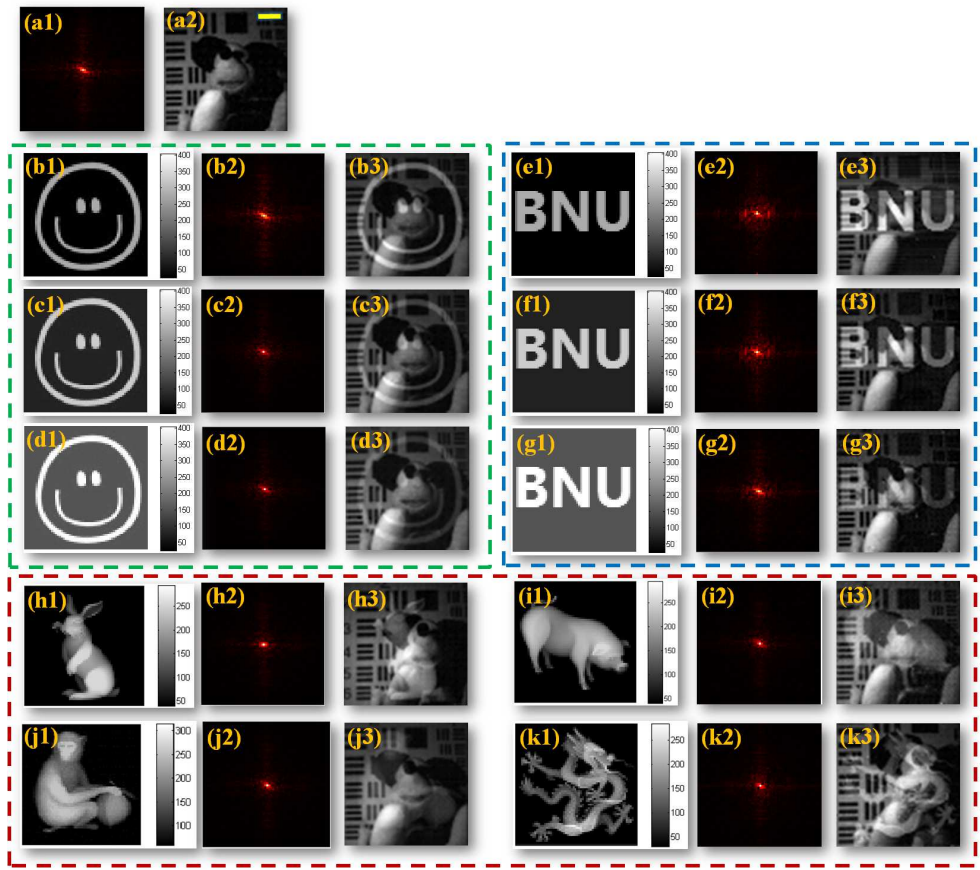}
\caption{Experimental results of TV-FSPI-based image fusion or visible watermark: (a1) the Fourier spectrum of the scene obtained by FSPI technique and (a2) the reconstructed image; (b1), (c1) and (d1) the watermark images of "smile" at different DC; (e1), (f1) and (g1) the watermark images of "BNU" at different DC; (h1)-(k1) the watermark images of complex grayscale animals; (b2)-(k2) the Fourier spectrum acquired in the experiments; (b3)-(k3) experimental results of single-pixel fusion of each watermark image.}
\end{figure}

In the experiment, the target we selected is a complex three-dimensional scene, in which the background board is the resolution board printed on the A3 paper (see Fig. 4(a2)). The size of the image to be reconstructed is 64$\times$64 pixels. According to the four-step FSPI technique, we need to play 16384 illumination patterns to obtain the spectrum information of the scene. In fact, because of the symmetry of the Fourier spectrum, we only need to play half of the illumination patterns, 8192 frames, i.e., the sampling rate is 200$\%$. Figs. 4(a1) and 4(a2) show the experimental results of the original FSPI technique without TV signals, where Fig. 4(a1) is the Fourier spectrum obtained by Eq. (\ref{eq:3}), and Fig. 4(a2) is the image reconstructed by Fig. 4(a1) using IFFT, in which the scale bar is 4 cm. Note that Fig. 4(a1) is pseudo-color processed and shows the real part of the logarithmic spectrum. In our optical experiments, we first verify the interesting inverse law and use this phenomenon to adjust the intensity of the watermark. We not only select simple binarized images as the watermarks (Figs. 4(b1)-4(g1)), but also select complex grayscale objects (Figs. 4(h1)-4(k1)). From the experimental results shown in Fig. 4, this scheme can realize the visual high-quality image fusion or watermarking under single-pixel detection.

\begin{figure}[!ht]
\centering
\includegraphics[width=13cm]{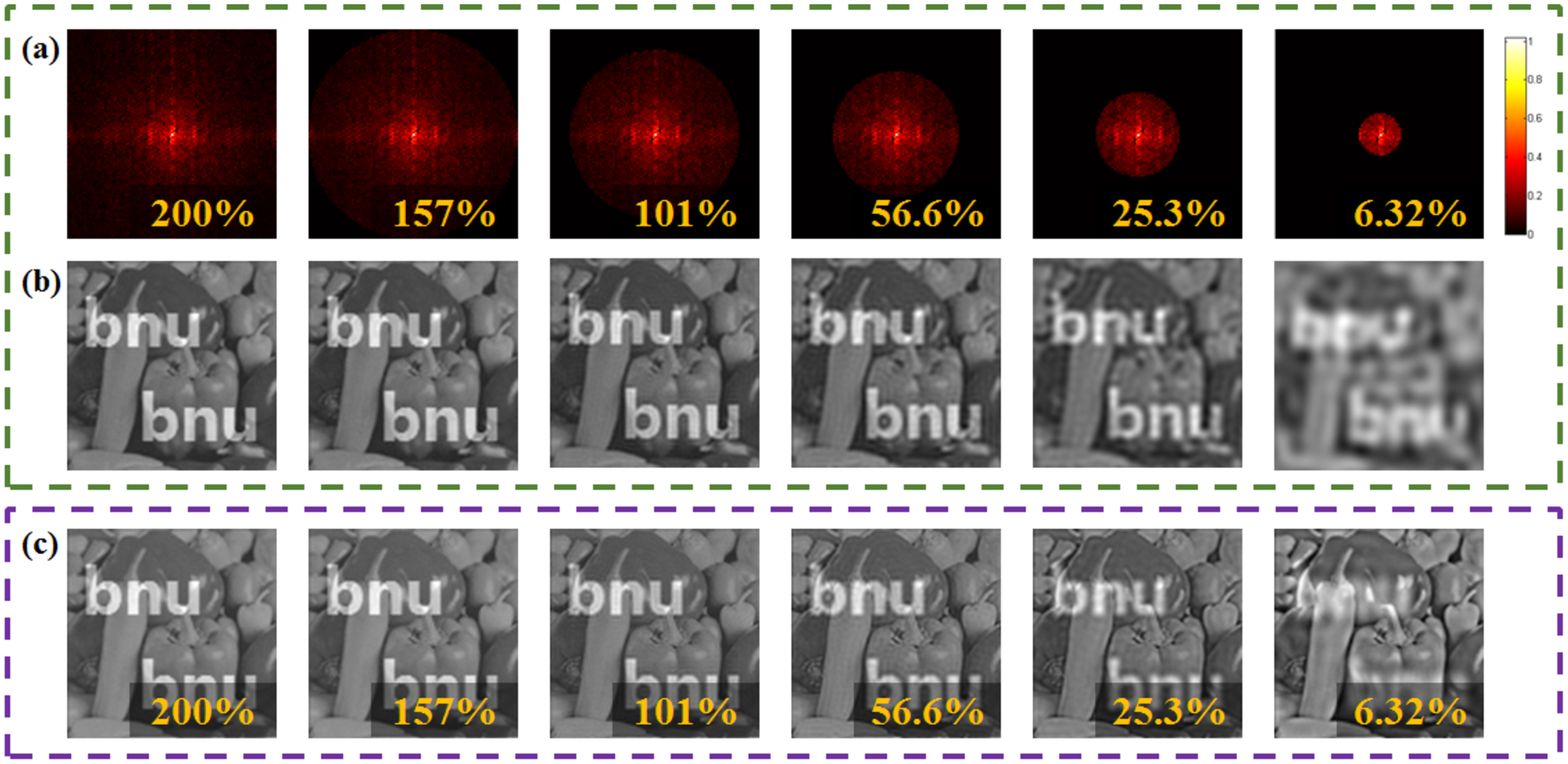}
\caption{Numerical simulations of compressive image watermarking using TV-FSPI scheme: (a) partially collected Fourier spectrum of the image, where the percentage of the yellow bold in the lower right corner is the sampling rate; (b) reconstructed image of (a); (c) image watermarking results when the watermark is compressively hidden in the low frequency region of the Fourier spectrum of the host image. The percentage of the yellow bold in the lower right corner is the ratio of the TV signal length to the number of samples. }
\end{figure}

\begin{figure}[!ht]
\centering
\includegraphics[width=13cm]{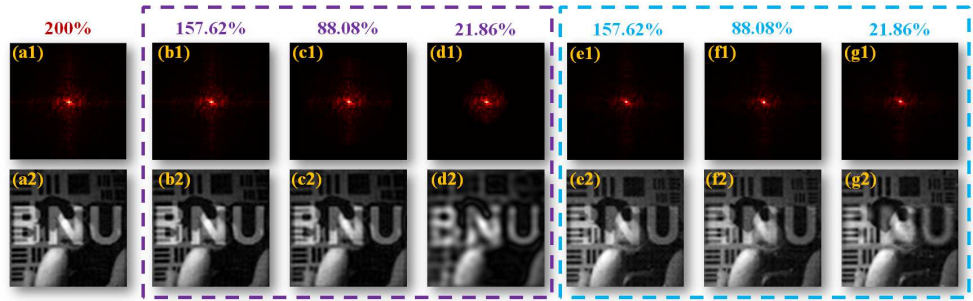}
\caption{Experimental results of compressive image watermarking using TV-FSPI scheme: (a1)-(d1) partial Fourier spectrum of the scene, where the percentage of the top purple bold is the sampling rate; (a2)-(d2) reconstructed image of (a1)-(d1); (e1)-(g1) the spectrum information collected at a sampling rate of 200$\%$, where the percentage of the uppermost sky blue bold indicates the ratio of the length of the TV signal to the number of samples; (e2)-(g2) image watermarking when the watermark is compressively hidden in the low frequency region of the Fourier spectrum of the host image.}
\end{figure}

\begin{figure}[!ht]
\centering
\includegraphics[width=13cm]{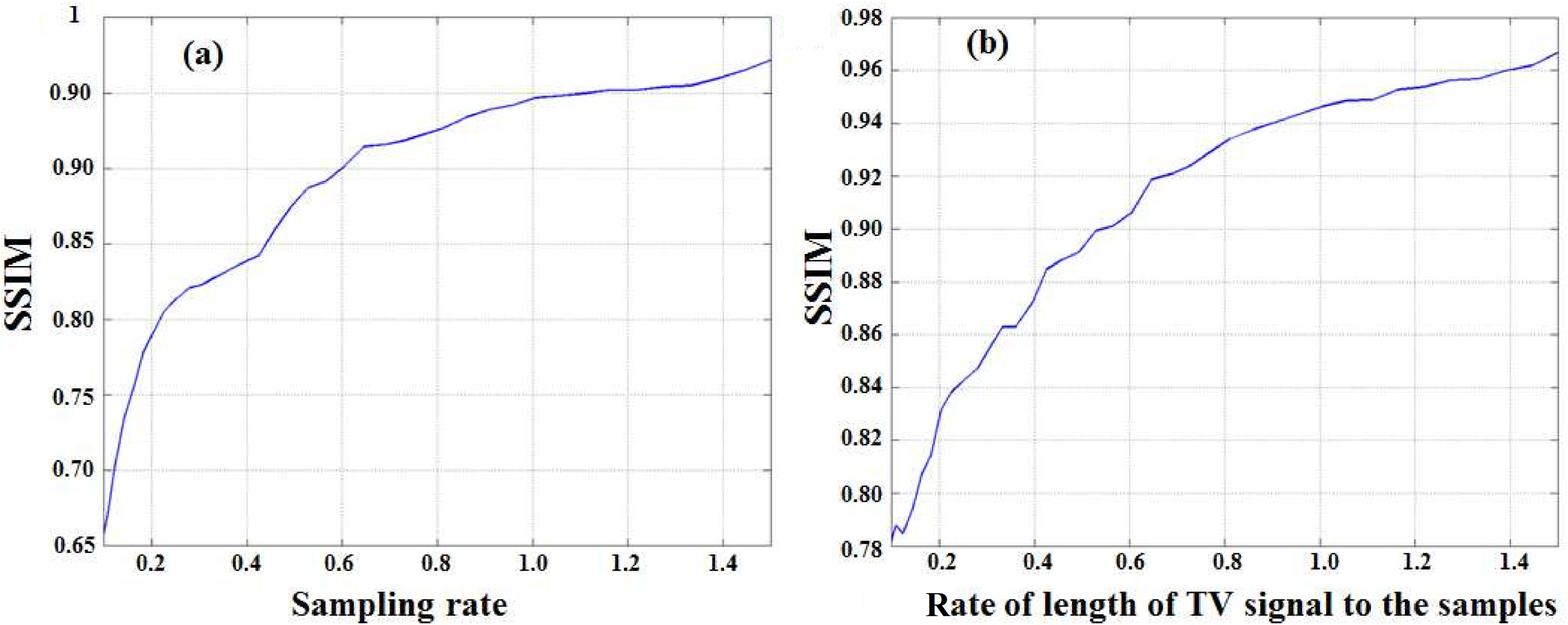}
\caption{Numerical quantitative analysis of the SSIM of the watermarked image with the sampling rate (a), and with the rate of the length of TV signal to the samples (b). }
\end{figure}

Since the FSPI technique applies measurement in the Fourier domain, where most natural objects are sparse, the large coefficients of natural objects are  concentrated at the low frequency region \cite{Zhang2015,Sen2009,P2009}. Therefore, the FSPI technique itself is highly compressible, that is, an image can be reconstructed adatively below the Nyquist sampling law. This scheme successfully inherits the advantages of FSPI technique in compressive imaging. It can complete single-pixel image fusion or watermarking under the condition of Nyquist sampling law and can greatly reduce the coefficients needed to realize image watermarking. Figs. 5 and 6 show the numerical simulations and experimental results of compressive image watermarking,  respectively. As shown in Figs. 5(a)-5(b) and Figs. 6(b)-6(d), we can only play the sinusoidal illumination patterns corresponding to the low frequency region to obtain the low frequency information of the scene and reconstruct the image by IFFT, which can greatly reduce the acquisition time and improve the imaging efficiency. On the other hand, we can achieve single-pixel image watermarking or fusion with a lower data volume on the premise of high sampling rate. As shown in Fig.5(c) and Figs.6(e)-6(g), their sampling rates are still set at 200$\%$. The difference is that we can greatly reduce the coefficients required by the watermark, which can reduce the length of the TV signal. For example, we can simply convert the low frequency coefficients of the watermark into TV signals that are hidden into the light source. Note that as the length of the TV signals become shorter, the host image is not distorted, and only the image quality of the watermark is gradually declining. In this way, we can use the shorter TV signals to complete the placement of the watermark information, which will be more practical and more concealed. As shown in Fig.7, we add the quantitative analysis of the fused image quality when the sampling rate changes. We select the watermarked image under the ideal full-sampling condition as the reference image and select SSIM as the evaluation index. Fig.7(a) corresponds to the setting of Figure 5(b), and Fig.7(b) corresponds to the setting of Fig.5(c).

\begin{figure}[!ht]
\centering
\includegraphics[width=13cm]{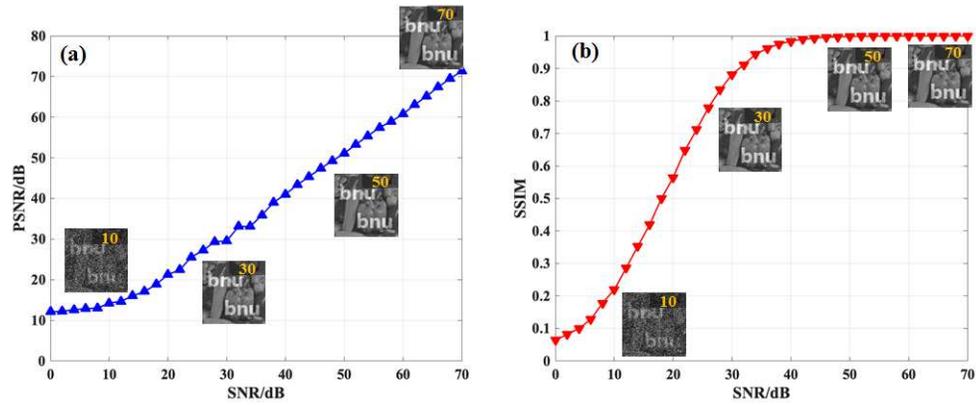}
\caption{Numerical simulation analysis of the noise immunity of TV-FSPI scheme: (a) the PSNR of the watermarked image with the change of the SNR of the system; (b) the SSIM of the watermarked image versus the SNR. The yellow number in the upper right corner of the images represents the corresponding SNR.}
\end{figure}

Next, we analyze the noise immunity of our scheme in the simulation. Since this scheme is a single-pixel acquisition to realize image watermarking, it is actually the encoding and decoding of TV signals, and it is mainly susceptible to Gaussian noise in this process. Therefore, we mainly consider the robustness of our scheme to Gaussian noise in the simulation. To quantify this problem, we select the watermarked image in the absence of noise as the reference image. As shown in Fig. 8, we plot the PSNR and SSIM curves with the change of the SNR of the system. It can be seen that when SNR of the system is greater than 30dB, the PSNR of the watermarked image is more than 30dB, and the SSIM of the watermarked image exceeds 0.8. The environment setting with SNR of 30dB is not difficult to implement in practice. Moreover, Eq. (\ref{eq:3}) shows that the differential detection adopted by this scheme can further improve the SNR of the system.

\subsection{De-watermarking}

\begin{figure}[!ht]
\centering
\includegraphics[width=13cm]{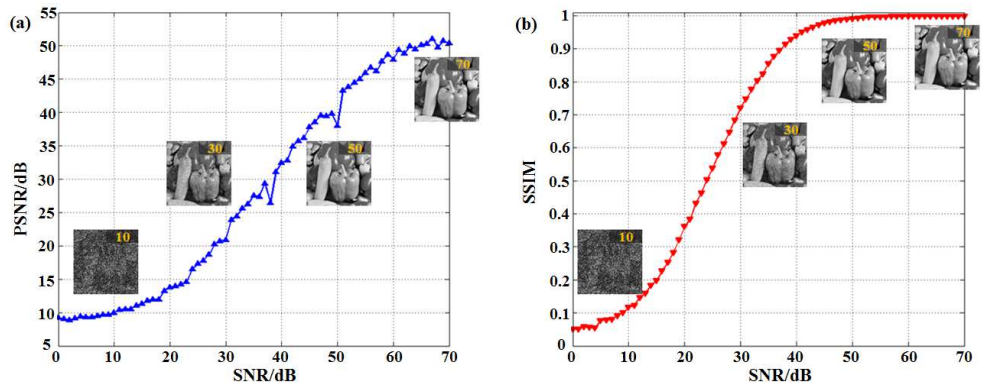}
\caption{Numerical simulation analysis of the noise immunity of the de-watermarking progress: (a) the PSNR of the watermarked image versus the SNR of the system; (b) the SSIM of the watermarked image versus the SNR. The yellow number in the upper right corner of the de-watermarked images represents the corresponding SNR.}
\end{figure}

\begin{figure}[!ht]
\centering
\includegraphics[width=13cm]{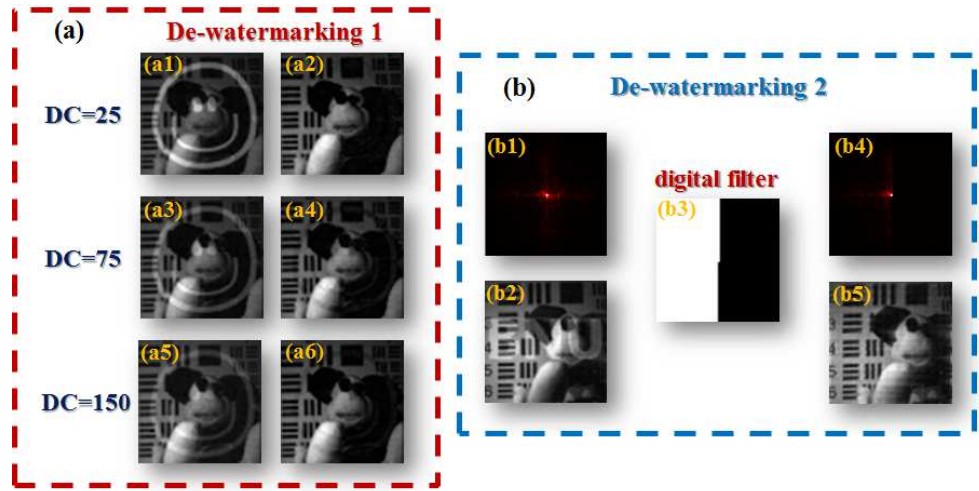}
\caption{Experimental results of two de-watermarking schemes: (a) original de-watermarking at different DC values; (b) de-watermarking by digital filtering.}
\end{figure}

The watermarking process of the TV-FSPI scheme at the transmitting end has been discussed from the simulations and experiments. Now we discuss the implementation of the de-watermarking at the receiving end. In order to remove the watermark, the transmitter must obtain the TV signals in advance. On the basis of the sequence of light intensity collected by the SPD, divided by the TV signal obtained in advance, the de-watermarked image can be reconstructed by Eqs.(3) and (4). As shown in Fig. 9, similar to the encoding process, we also analyze the noise immunity of the de-watermarking process in the simulation. However, encoding and decoding are a continuous process, which means that the noise is additive. Therefore, when the SNR of the system is 30dB, the PSNR of the de-watermarked image is only 20dB, and the SSIM is only 0.7. At the same time, when the SNR is too low, the image quality will decline rapidly, and the residual shadow of watermark will still appear in the image.

As shown in Fig. 10(a), we performed three de-watermarking experiments with different DC background terms. Although we can achieve high-quality de-watermarking results, there are still weak residual image of the watermark. We think the reason is that the TV signal is not completely multiplicative in the illumination mode, which is the inevitable systematic error of such decoding. To overcome the inherent limitations of this de-watermarking scheme, we develope a novel FSPI-based watermarking and de-watermarking scheme. In the previous section we mentioned that the watermark can be hidden in the low frequency region of the host image to shorten the length of the TV signal. Here, considering the symmetry of the Fourier spectrum, we can hide the watermark in half of the spectrum area of the host image or in certain areas. As shown in Fig. 10(b), we perform a simple supplemental experiment to illustrate the feasibility of this strategy. Half of the spectral information of the watermark is written into the light source, and the watermarked image can still be obtained after single-pixel detection. Next, we can implement de-watermarking by using a simple digital filter to erase the spectral region containing the watermark's information. The advantage of this de-watermarking scheme is that the watermark can be perfectly eliminated without leaving a trace of residual information. Of course, the user can choose the spectral region in which the watermark is carried, not only in the low frequency region or in the half of the spectral region as we proposed. Therefore, the digital filter itself can be regarded as a key to dewatermarking, which will further improve the practicability of the scheme.
\subsection{Encrypted watermarking}

\begin{figure}[!ht]
\centering
\includegraphics[width=11cm]{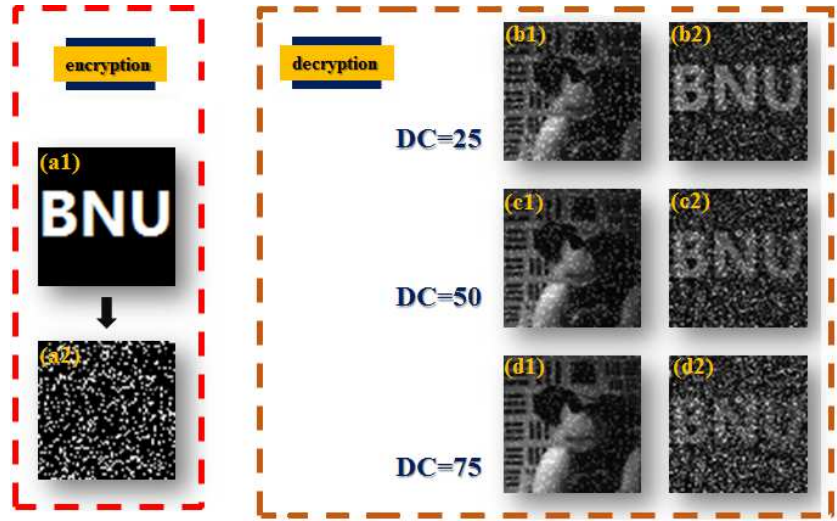}
\caption{Experimental results of encrypted FSPI-based single-pixel image watermarking: (a1) original watermark pattern; (a2) encrypted watermark pattern; (b1)-(d1)  image watermarking under different background DC values; (b2)-(d2) watermark extraction.}
\end{figure}

If the transmitter wants to secretly transmit the watermark pattern, we also provide a solution. As shown in Figure 11, we scramble the spatial information of the watermark pattern in a random way, and the encrypted watermark is converted into a TV sequence by FSPI and hidden into the light source. This random mapping is equivalent to the key to extract the watermark, which is only owned by the transmitter and the designated receiver. We did three watermark extraction experiments with different DC background terms, where the watermark pattern was written into the host image like speckle noise. This application will have an inherent background item, which is an image of the randomly mapped host scene, so we can adjust the intensity of the DC component to adjust the visibility of the watermark extraction. The larger the value of DC is, the worse the visibility of watermark extraction is, but the higher the concealment is in the watermarked image. As the transmitter, the value of DC can be flexibly adjusted to achieve the most suitable result. At the same time, this scheme has the potential to be combined with other steganographic techniques \cite{Liansheng2018,Zhang2019}.
\subsection{Full-color visible watermarking}

\begin{figure}[!ht]
\centering
\includegraphics[width=13cm]{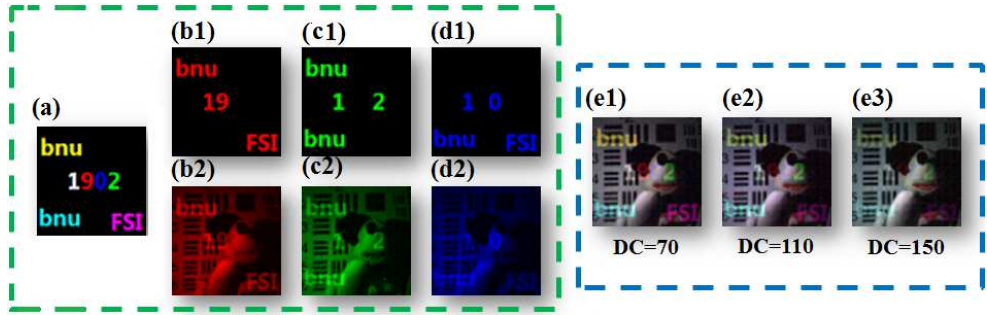}
\caption{Experimental results of FSPI-based full-color visible watermarking: (a) the colored watermark; (b1) the red (R) channel, (c1) the green (G), and (d1) the blue (B) channel of colorful watermark pattern; (b2)-(d2) SPI watermarking of RGB channels; (e1)-(e3) full-colored image watermarking under different values of DC background.}
\end{figure}

SPI has been applied in the field of full-color compressive imaging \cite{Welsh2013} and even multi-spectral imaging \cite{Li2017}. As shown in Fig.12, we present three complementary experimental results for single-pixel full-color image watermarking. Fig.12(a) shows the colored watermark, and we separate the colored watermark into three channels of RGB and generate corresponding TV signals. Then three sets of TV signals are loaded on the red, green, and blue illumination sequences, which is not difficult for color commercial projectors. In this way, we obtain the spectral information of the three channels of the scene through three single-pixel acquisitions to reconstruct the color watermarked image. Similar to the single-channel image watermark, we can still adjust the visibility of the color watermark in the host image by adjusting the DC background of each channel of the color watermark. We acknowledge the inevitable color distortion in the imaging results, which may be due to the SPD's unit response to the three colors. It is foreseeable that we may be able to use this scheme to place high-dimensional information in single-pixel multi-spectral imaging to achieve higher levels of information interaction and encryption.

\section{Extension to invisible watermarking based on TV-FSPI technique}

The invisible image watermarking or steganography plays a vital role in the field of information security. Here, we also propose a novel invisible watermarking scheme based on the TV-FSPI technique. In fact, we can consider this scheme as a single-pixel steganography only using weighed light source. Unlike previous frameworks of SPI system, there is no need to re-encode the illumination mode and no extxa complex optimization algorithms are required.

\begin{figure}[!ht]
\centering
\includegraphics[width=14cm]{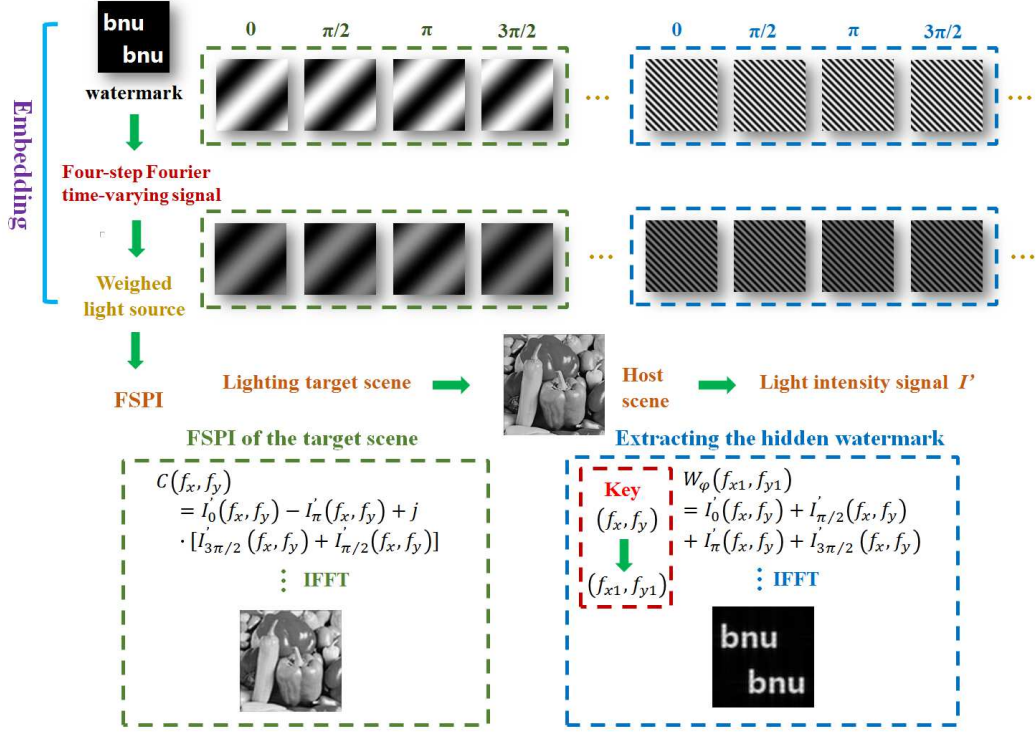}
\caption{The schematic diagram of the steganography scheme based on the TV-FSPI scheme.}
\end{figure}

Notably, we still need to meet the limits proposed in the framework of TV-FSPI technique, that is, do not recode the sinusoidal illumination mode, and simply modify the weighting coefficient of the light source. Therefore, this steganography scheme still has a high concealment characteristic because it secretly embeds information in the light source at the transmitting end. Fig.13 shows the schematic diagram of the steganography scheme. Much different from the visible optical watermarking technique proposed in the Section 3, the core of the steganography scheme is to hide the spectral information of the watermark in the Fourier spectrum of the host scene to be acquired. The acquisition method is the same as that used in the Section 3, requiring four frames of sinusoidal illumination patterns acquisition for a single spatial frequency component. However, the four coefficients corresponding to the four-frame illumination patterns, share the same value, where this value represents one of the four-step coefficients of the invisible watermark. In this way, the one-to-one correspondence between this type of watermark image and the host image in the Section 3 no longer holds, and the host image in the spatial domain disappears after the IFFT is calculated using Eqs. (\ref{eq:3})  and  (\ref{eq:4}). In other words, the light intensity value carrying the host scene information carries the information in the watermark image at the same time. Since Eq. (11) is established, we can extract the time-varying coefficients of the watermark losslessly by the following equation:

\begin{equation}
W_{\varphi}(f_{x1},f_{y1})=I'_{0}(f_{x},f_{y})+I'_{\pi/2}(f_{x},f_{y})+I'_{\pi}(f_{x},f_{y})+I'_{3\pi/2}(f_{x},f_{y}).
\label{eq:19}
\end{equation}

In theory, four-step time-varying coefficients (i.e., a single Fourier coefficient using Eq. (\ref{eq:6})) of the watermark can be obtained for every sixteen frames of sinusoidal illumination patterns, which means the maximum data capacity of the steganography scheme is one quarter of the visible watermarking technique in the Section 3. Therefore, the size of the watermark should be smaller than the size of the host scene, or we can embed only the low frequency information of the watermark in the time-varying light source. Although the data capacity is limited, from another aspect, the steganography scheme is more robust to noise. The reason is that multiple summation is actually a relatively average process. Meanwhile, this process can reduce the dependence on the modulation deep of the light source. If we load the value of the Fourier coefficient directly (in two steps, the real part and the imaginary part), in theory this can actually increase the capacity by twice, but it requires a much higher modulation depth because the fluctuation range of the four-step coefficients are far less than that of the Fourier coefficients in two steps.

\begin{figure}[!ht]
\centering
\includegraphics[width=12cm]{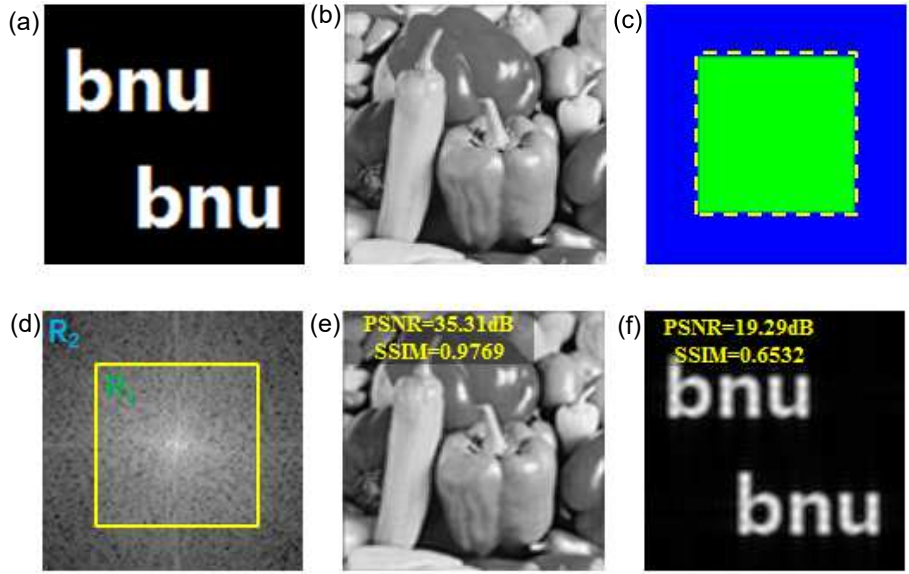}
\caption{A numerical simulation example using the steganography scheme: (a) watermark; (b) host scene; (c) mask of the Fourier frequency domain; (d) numerically simulated Fourier spectrum; (e) the reconstructed host scene; (f) the extracted watermark.}
\end{figure}

As shown in Fig.14, we demonstrate a numerical simulation example using the scheme. Fig.14(a) is the watermark (100$\times$100 pixels) to be embedded, and Fig.14(b) is the host image (100$\times$100 pixels). Fig.14(c) is the mask of the Fourier frequency domain, where the green area (60$\times$60 pixels) represents the Fourier spectral region of the target scene (no watermark embedding) and the blue region represents the Fourier spectral region with watermark embedding. The purpose of our design is to minimize the interference of the watermark embedding on the host scene. Of course, users can design arbitrary masks and scales to weigh the ratio of the target scene to the watermark in the Fourier spectral region. Thus, the amount of information left to the embedded watermark information is only 6400 pixels. The acquisition process of each pixel value can extract one four-step coefficient of a watermark using Eq.(\ref{eq:19}), which implies that the actual size left for the watermark in the Fourier spectral region is only 1600 pixels. In order to reconstruct a watermark image of size 100$\times$100 pixels, 1600 Fourier coefficients of low frequency region make this compressive reconstruction possible. Fig.14(d) shows the simulated Fourier spectrum, where the region of $R_{1}$ represents the low-frequency region without the watermark embedded, and the presence of this region ensures high fidelity of the target scene (the PSNR reaches 35dB and the SSIM is close to 0.98 shown in Fig.14(e)). The region of $R_2$ is multiplexed with two parts of information, including the high frequency area of the host scene and the low frequency area of the watermark. Using Eq.(\ref{eq:19}), the watermark hidden in the host scene can be extracted, as shown in Fig.14(f). The PSNR of the extracted watermark is about 20 dB, and the SSIM is about 0.65. This is actually a process of under-sampling reconstruction (effective data is only 16\% of the original watermark). Then the user can set the appropriate sampling rate by adjusting the design of the relative ratio of the mask in Fig.14(c). Of course, we can also reduce the size of the watermark. The relationship between the two is not discussed in detail here. On the other hand, in order to illustrate the invisibility of watermark in the reconstruted host scene of the spatial domain, we calculate the SSIM value between the reconstructed host scene Fig.14(e) and the watermark pattern Fig.14(a). The SSIM value goes to -0.0021, which means that the correlation between the two is extremely low or even unrelated. This is also in line with the prediction of the above theory that the unordered one-to-four multiplexing of the Fourier domain is unlikely to visualize the watermark in the spatial domain.

Meanwhile, the region of $R_2$ must contain both the host image and the watermark due to the inherent characteristics of the TV-FSPI technique, but the information is expressed or decoded differently. So what is the impact of the presence of the $R_2$ region on the image quality of the host scene? Fig.15 shows the reconstructed images of the host scene using different regions, where Fig.15(a) is the ground truth, Fig.15(b) is the reconstructed image of the host scene using all regions, and Fig.15(c) is the reconstructed image using only region $R_1$. The numerical results suggest that although there is watermark embedding in the region of $R_2$, it still contributes to the improvement of the PSNR and resolution of the reconstructed host scene.

\begin{figure}[!ht]
\centering
\includegraphics[width=12cm]{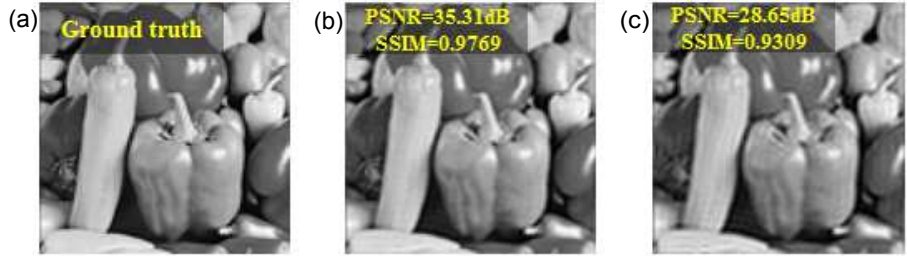}
\caption{The reconstructed images of the host scene using different regions in Fig.14(d): (a) ground truth; (b) reconstructed image of the host scene using all regions, $R_{1}$ and $R_{2}$; (c) reconstructed image using only region $R_{1}$.}
\end{figure}

Next we discuss the security performance of this steganography scheme based on the TV-FSPI technique. There are two aspects that can be used as the keys to extract the watermark at the same time, one is the mapping relations between ($f_x$,$f_y$) and ($f_{x1}$,$f_{y1}$) according to Eq. (\ref{eq:19}), and the other is the design of mask shown in Fig.14(c). Both can be highly random, and it is impossible for users without prior knowledge to extract the invisble watermark. In comparison, the former is more flexible to use, while the latter may be subject to some limitations, such as trying not to multiplex watermark information in the low frequency region of the host scene.

Finally, we discuss the noise immunity of steganography scheme. In SPI, we collect time-varying light intensity signals, so we mainly consider the influence of Gaussian noise on this steganography scheme, which is shown in Fig.16. Figs.16(a)-16(b) show the PSNR and the SSIM of the reconstructed host scene and the extracted watermark with the change of the SNR of the system, respectively. Similar to what is previously conceived, the extracted watermark has higher noise immunity than the reconstructed host scene due to four summations. Even when the reconstructed image of the scene is completely submerged by the noise (i.e., SNR=0 dB), we can still successfully extract the watermark with a not bad quality (see Figs.16(c)-16(d)).

\begin{figure}[!ht]
\centering
\includegraphics[width=13cm]{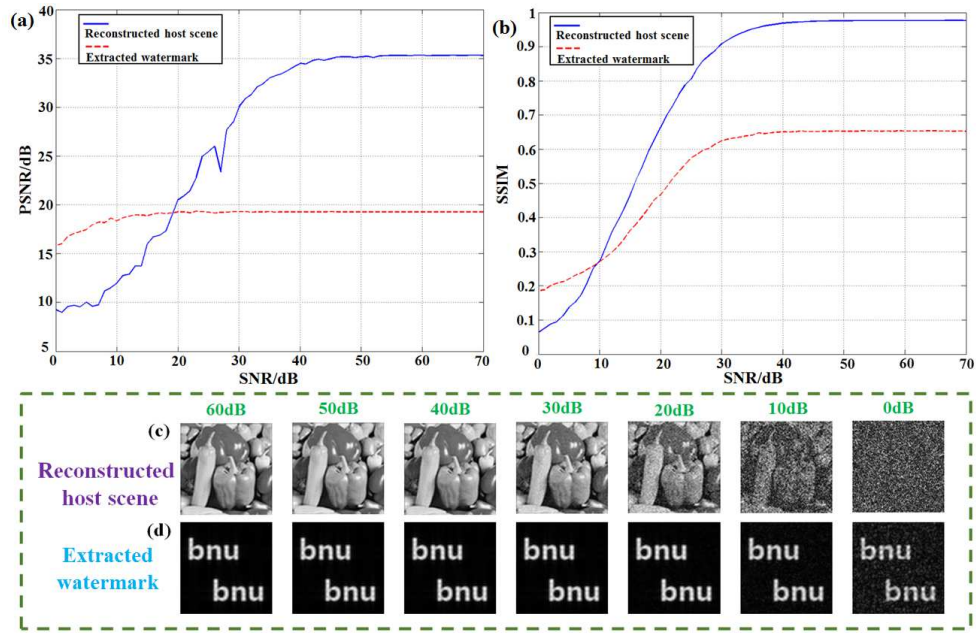}
\caption{Numerical simulation analysis of the noise immunity of the TV-FSPI-based steganography scheme: (a) the PSNR of the reconstructed host scene and the extracted watermark with the change of the SNR of the system; (b) the SSIM of the reconstructed host scene and the extracted watermark versus the SNR; (c) the reconstructed images of host scene at different SNRs; (d) the images of extracted watermark at different SNRs.}
\end{figure}

\section{Discussion and conclusion}

\begin{figure}[!ht]
\centering
\includegraphics[width=13cm]{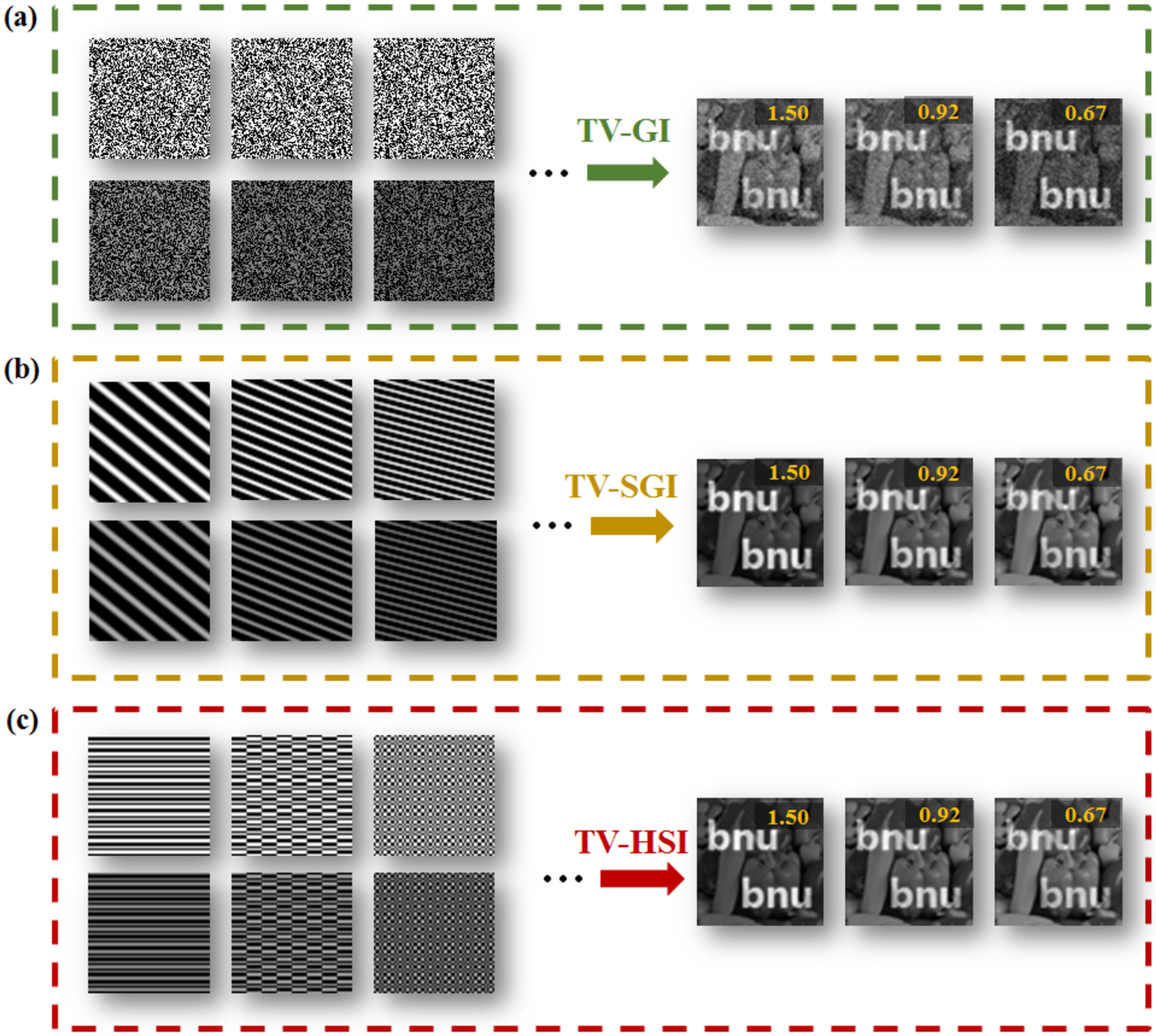}
\caption{Simulation results of image watermarking in other illumination modes: (a) random illumination mode; (b) sinusoidal orthogonal illumination mode; (c) differential Hadamard orthogonal illumination mode.}
\end{figure}

The performance and application in image fusion in spatial domain or watermarking based on TV-FSPI technique are fully illustrated in the simulations and experiments in the Sections 3 and 4. The introduction of multiplicative TV signals will add an unexpected degree of freedom to SPI, which is worthy of our exploration. In addition to introducing TV signals into the light source, the sinusoidal illumination mode is adopted based on FSPI technique. Although this imaging method has its unique advantages, this mode also has a drawback, that is, it is limited by the modulation depth of the modulation system, and the modulation speed is not very fast. Are other modulation modes, such as binary Hadamard illumination mode \cite{Sun2017,Zhang2017} or binary random illumination mode, compatible with this scheme? The answer is yes. As shown in Fig. 17, we find in the simulations that the multiplexing of TV signals can not only achieve image watermarking under the four-step FSPI framework, but also achieve the same function in the frameworks of three-step FSPI, traditional CGI (random illumination mode), sinusoidal ghost imaging (sinusoidal orthogonal illumination mode) \cite{Khamoushi2015} and Hadamard SPI (differential Hadamard orthogonal illumination mode). Meanwhile, these schemes also satisfy the inverse law similar to TV-FSPI scheme, so we can adjust the visibility of watermark flexibly by changing the value of $Q$, which also indicates the high extensibility and compatibility of this scheme. Our image watermarking scheme can perfectly match the recently proposed single-pixel broadcast system \cite{Zhong2018}. Firstly, our scheme is applicable to a wide range of illumination modes. Secondly, instead of re-coding the illumination modes, we only need to hide the TV signal in the light source of the transmitting end, which is highly concealed. Thirdly, our scheme can achieve adaptive compressive sampling, and can realize encrypted image watermarking, full-colored image watermarking and even multi-spectral high-dimensional watermarking. At the same time, the scheme allows the length of the TV signal to be much lower than the Nyquist sampling law.

In summary, we have proposed the novel TV-FSPI watermarking scheme, confirmed through theory, simulation and experimentation. The TV signal is hidden in the light source of the SPI system to achieve high-quality compressive image watermarking or image fusion, which has high robustness for not only simple binary images, but also complex grayscale images and even color images. Moreover, our scheme has high scalability and is compatible with the existing mainstream illumination modes, which has the potential to be applied in single-pixel broadcasting system and multi-spectral single-pixel imaging system. We believe that our scheme could pave the way for the application of active SPI in visible and invisible image watermarking and public information security.

\section*{Funding}
 National Natural Science Foundation of China (11474027, 11735005, 61675028), and the Interdiscipline Research Funds of Beijing Normal University.

\bibliographystyle{}

\begin{thebibliography}{99}
\bibitem{Shapiro2008}J. H. Shapiro, "Computational Ghost Imaging," Physical Review A \textbf{78},061802 (2008).
\bibitem{Bromberg2008}Y. Bromberg,  O. Katz, and Y. Silberberg, "Ghost imaging with a single detector," Physical Review A \textbf{79},1744-1747 (2008).
\bibitem{Sun2013}B. Sun, M. P. Edgar,  R. Bowman, L. E. Vittert,  S. Welsh,  A. Bowman, and M. J. Padgett, "3D computational imaging with single-pixel detectors," Science \textbf{340}, 844-847 (2013).
\bibitem{Sun2016}M. J. Sun,  M. P. Edgar,  G. M. Gibson,  B. Sun,  N. Radwell, R. Lamb, and M. J. Padgett, "Single-pixel three-dimensional imaging with time-based depth resolution," Nature Communications, \textbf{7},12010 (2016).
\bibitem{Soldevila2018}F. Soldevila, V. Durán, P. Clemente, J. Lancis, and E. Tajahuerce, "Phase imaging by spatial wavefront sampling," Optica  \textbf{5},164-174 (2018).
\bibitem{Hu2019}X. Hu, H. Zhang, Q. Zhao, P. Yu, Y.  Li, and L. Gong, "Single-pixel phase imaging by Fourier spectrum sampling," Applied Physics Letters \textbf{114}, 051102 (2019).
\bibitem{Liu2019}R. Liu, S. Zhao, P. Zhang, H. Gao, and F. Li, "Complex wavefront reconstruction with single-pixel detector," Applied Physics Letters \textbf{114}, 161901 (2019).
\bibitem{Radwell2014}N. Radwell, K. J. Mitchell, G. M. Gibson, M. P. Edgar, R. Bowman, and M. J. Padgett, "Single-pixel infrared and visible microscope," Optica \textbf{1}, 285 (2014).
\bibitem{Stantchev2016}R. I. Stantchev, B. Sun, S. M. Hornett, P. A. Hobson, G. M. Gibson, M. J. Padgett, and E. Hendry, "Noninvasive, near-field terahertz imaging of hidden objects using a single-pixel detector," Science Advances \textbf{2}, e1600190 (2016).
\bibitem{Edgar2019}M. P. Edgar, G. M. Gibson, and M. J. Padgett, "Principles and prospects for single-pixel imaging," Nature Photonics \textbf{13}, 13-20 (2019)
\bibitem{Ferri2010}F. Ferri, D. Magatti, L. A.  Lugiato, and A. Gatti, "Differential ghost imaging," Phys. Rev. Lett. \textbf{104}, 219902 (2010).
\bibitem{Sun2012}B. Sun, S. S. Welsh, M. P. Edgar, J. H. Shapiro, and M. J. Padgett, "Normalized ghost imaging," Opt. Express \textbf(20), 16892-16901 (2012).
\bibitem{Khamoushi2015}S. M. Khamoushi, Y. Nosrati, and S. H. Tavassoli, "Sinusoidal ghost imaging," Optics Letters \textbf{40}, 3452-3455 (2015).
\bibitem{Li2013}M. F. Li, Y. R. Zhang, X. F. Liu, X. R. Yao, K. H. Luo, H. Fan, and L. A. Wu, "A double-threshold technique for fast time-correspondence imaging," Applied Physics Letters \textbf{103}, 211119 (2013).
\bibitem{Zhang2014}C. Zhang, S. Guo, J. Cao, J. Guan, and F. Gao, "Object reconstitution using pseudo-inverse for ghost imaging," Opt. Express,  \textbf{22}, 30063 (2014).
\bibitem{Gong2015}W. Gong, "High-resolution pseudo-inverse ghost imaging," Photonics Research, \textbf{3}, 234 (2015).
\bibitem{Duarte2008}M. F. Duarte, M. A. Davenport, D. Takhar, J. N. Laska, T. Sun, K. F. Kelly, and R. G. Baraniuk, "Single-pixel imaging via compressive sampling," \emph{IEEE signal processing magazine} \textbf{25}, 83-91 (2008).
\bibitem{Zhang2015}Z. Zhang, X. Ma, and J. Zhong, "Single-pixel imaging by means of fourier spectrum acquisition," Nature Communication \textbf{6}, 6225 (2015).
\bibitem{Zibang2018}Z. Zhang, S. Liu, J. Peng, M. Yao, G. Zheng, and J. Zhong, "Simultaneous spatial, spectral, and 3d compressive imaging via efficient fourier single-pixel measurements," Optica \textbf{5}, 315-319 (2018).
\bibitem{Huang2018}J. Huang, D. Shi, K. Yuan, S. Hu, and Y. Wang, "Computational-weighted Fourier single-pixel imaging via binary illumination," Opt. Express \textbf{26}, 16475-16559 (2018).
\bibitem{Jiang2018}H. Jiang, H. Liu, X. Li, and H. Zhao, "Efficient regional single-pixel imaging for multiple objects based on projective reconstruction theorem," Optics and Lasers in Engineering \textbf{110}, 33-40 (2018).
\bibitem{Czajkowski2018}K. M. Czajkowski,  A. Pastuszczak, and R. Koty\'{n}ski, "Real-time single-pixel video imaging with Fourier domain regularization," Opt. Express \textbf{26}, 20009-20022 (2018).
\bibitem{Li1995}H. Li, B. S. Manjunath, and S. K. Mitra, "Multi-Sensor Image Fusion using the Wavelet Transform," \emph{Graphical models and image processing} \textbf{57}, 235-245 (1995).
\bibitem{Cox2002}I. J. Cox, M. L. Miller, J. A. Bloom, and C. Honsinger, "Digital watermarking," San Francisco: Morgan Kaufmann \textbf{53} (2002).
\bibitem{Hu2005}Yongjian Hu, Sam Kwong, and Jiwu Huang. "An algorithm for removable visible watermarking." \emph{IEEE Transactions on Circuits and Systems for Video Technology} \textbf{16} 129-133 (2005).
\bibitem{Thodi2007}D. M. Thodi, and J. J. Rodr\'{\i}guez, "Expansion embedding techniques for reversible watermarking," \emph{IEEE Transactions on Image Processing} \textbf{16}, 721-730 (2007).
\bibitem{Xu2016}Wen-Hui Xu, Hong-Feng Xu, Yong Luo, Tuo Li, and Yi-Shi Shi, "Optical watermarking based on single-shot-ptychography encoding," Opt. Express \textbf{24}, 27922-27936 (2016).
\bibitem{Yang2018}Na Yang, Qian-Kun Gao, and Yi-Shi Shi, "Visual-cryptographic image hiding with holographic optical elements," Opt. Express \textbf{26}, 31995-32006 (2018).
\bibitem{Chen2014}W. Chen, and X. Chen, "Marked ghost imaging," Applied Physics Letters \textbf{104}, 251109 (2014).
\bibitem{Li2012}H. Li, Z. Chen, J. Xiong, and G. Zeng, "Periodic diffraction correlation imaging without a beam-splitter," Optics Express \textbf{20}, 2956-2966 (2012).
\bibitem{Sunm2013}M. Sun, J. Shi, H. Li, and G. Zeng, "A simple optical encryption based on shape merging technique in periodic diffraction correlation imaging," Opt. Express \textbf{21}, 19395-19400 (2013).
\bibitem{Dongfeng2017}S. Dongfeng, H. Jian, W. Yingjian, Y. Kee, X. Chenbo, L. Dong, and Z. Wenyue, "Simultaneous fusion, imaging and encryption of multiple objects using a single-pixel detector," Scientific Reports \textbf{7}, 13172 (2017).
\bibitem{Liansheng2018}S. Liansheng, C. Yin, T. Ailing, and A. K. Asundi, "An optical watermarking scheme with two-layer framework based on computational ghost imaging," Optics and Lasers in Engineering \textbf{107}, 38-45 (2018)
\bibitem{Zhang2019}C. Zhang, W. He, B. Han, M. Liao, D. Lu, X. Peng, and C. Xu, "Compressive optical steganography via single-pixel imaging," Opt. Express \textbf{27}, 13469-13478 (2019).
\bibitem{Zhong2018}J. Zhong, M. Yao, S. Jiao, L. Xiang, and Z. Zhang, "Secured single-pixel broadcast imaging," Opt. Express, \textbf{26}, 14578-14591 (2018).
\bibitem{SunM2016}M. J. Sun, M. P. Edgar, and D. B. Phillips, G. M. Gibson, and M. J. Padgett, ``Improving the signal-to-noise ratio of single-pixel imaging using digital microscanning,'' Opt. Express, \textbf{24}, 10476-10485 (2016).
\bibitem{Alain2010}A. Hore, and D. Ziou, "Image quality metrics: PSNR vs. SSIM," In 2010 20th International Conference on Pattern Recognition 2366-2369 (2010).
\bibitem{Sen2009}P.  Sen, and S. Darabi,  "Compressive dual photography. Comput," Graph. Forum \textbf{28}, 609-618 (2009).
\bibitem{P2009}P.  Peers, et al. "Compressive light transport sensing," ACM Trans. Graph. \textbf{28}, 3 (2009).
\bibitem{Welsh2013}S. S. Welsh, M. P. Edgar, R. Bowman, P. Jonathan, B. Sun, and M. J. Padgett, "Fast full-color  computational imaging with single-pixel detectors," Optics Express \textbf{21}, 23068-23074 (2013).
\bibitem{Li2017}Z. Li, J. Suo, X. Hu, C. Deng, J. Fan, and Q. Dai, "Efficient single-pixel multispectral imaging via non-mechanical spatio-spectral modulation," Scientific Reports \textbf{7}, 41435(2017).
\bibitem{Sun2017}M. J. Sun, L. T. Meng, M. P. Edgar, M. J. Padgett, and N. Radwell, "A russian dolls ordering of the hadamard basis for compressive single-pixel imaging," Scientific Reports \textbf{7}, 3464 (2017).
\bibitem{Zhang2017}Z. Zhang, X. Wang, G. Zheng, and J. Zhong, "Hadamard single-pixel imaging versus Fourier single-pixel imaging," Opt. Express  \textbf{25}, 19619-19639 (2017).


\end{thebibliography}

\end{document}